\documentclass[twocolumn,preprintnumbers,amsmath,amssymb]{revtex4}
\usepackage{graphicx}
\usepackage{dcolumn}
\usepackage{bm}
\begin{document}
%
%\preprint{APS/123-QED}
%
\title
	{Susceptibility of a Magnetic Impurity \\
	in Two-Dimensional Disordered Electron Systems}
\author
	{Takuma Ohashi and Sei-ichiro Suga}
\affiliation
	{Department of Applied Physics, Osaka University
	, Suita, Osaka 565-0871}
\date{\today}% It is always \today, today,
\begin{abstract}
We investigate the Kondo effect in two-dimensional disordered electron systems using a finite-temperature quantum Monte Carlo method.  Depending on the position of a magnetic impurity, the local moment is screened or unscreened by the spin of the conduction electron. The results suggest that the Kondo temperature takes different values depending on the position of a magnetic impurity. We show that the distribution of the Kondo temperature becomes wide and the weight at $T_K=0$ becomes large as randomness increases. 
The average susceptibility shows a weak power-law divergence at low temperature in strongly disordered systems, indicating a non-Fermi-liquid behavior. 
We calculate the correlation function between the local moment and the spin of the conduction electron. The results are discussed in connection with Kondo screening. 
\end{abstract}
%                             % Classification Scheme.
\maketitle
%%%%%%%%%%%%%%%%%%%%%%%%%%%%%%%%%%%%%%%%%%%%%%%%%%%%%%%%%%%%%%%%%%%%
%           Main text
%%%%%%%%%%%%%%%%%%%%%%%%%%%%%%%%%%%%%%%%%%%%%%%%%%%%%%%%%%%%%%%%%%%%
%
In low-dimensional disordered systems, the conduction electron localizes within a characteristic length called the localization length \cite{And,AALR}. It was shown that the localization effect enhances the electron correlation effects significantly, leading to conspicuous features particularly in low-dimensional systems \cite{AA,HF,LR,Fink,DiC,BK}. The Kondo effect is a typical phenomenon caused by the electron correlation around a magnetic impurity.
The Kondo effect in disordered systems was investigated using the perturbation expansion in the weakly localized regime \cite{Ohkawa,VZ,Suga1,Suga2}. 
On the basis of the $s-d$ model with random potentials, Ohkawa {\it et al.} \cite{Ohkawa} showed that the Kondo logarithmic terms for conductivity and susceptibility are modified into anomalous terms. 
Within the most divergent approximation, it was shown that higher order corrections consist of the product of such anomalous terms and the Kondo logarithmic terms, and that the latter is scaled into the same Kondo temperature as that in an ordinary Kondo system \cite{Suga1}. 
The temperature dependence of the conductivity and the susceptibility around and below the Kondo temperature in the weakly localized regime was calculated using the $1/N$ expansion based on the Coqblin-Schrieffer model \cite{Suga1} and the perturbation expansion based on the Anderson model \cite{Suga2}, respectively.

The Kondo effect in strongly disordered systems was studied by Dobrosavljevi\'c {\it et al.} by taking account of the Coulomb interaction among conduction electrons \cite{DKK}.  On the basis of a slave-boson approach, they showed that, as a result of the localization and interaction effects, the Kondo temperature is modified, and that the magnetic susceptibility and the coefficient of the linear specific heat show divergences as temperature approaches zero. 
In fact, a strong spatial distribution in the Kondo temperature was suggested from experimental results of strong broadening of the Cu NMR line of ${\rm UCu_{5-x}Pd_x}$ \cite{NMR}. 
In spite of these findings, the effects of strong randomness itself on the behavior of a magnetic impurity have not yet been fully investigated from a microscopic viewpoint.

In this letter, we study the Kondo effect in two-dimensional (2D) strongly disordered electron systems. Using a finite-temperature quantum Monte Carlo (QMC) method \cite{Hirsch,GHS}, we calculate the susceptibility of a magnetic impurity and the correlation function between the local moment and the spin of the conduction electron.

Let us consider the single-impurity Anderson model with on-site random potentials described by the Hamiltonian 
%%%%%%%%%%%%%%%%%%%%%%%%%%%%%%%%%%%%%%%%%%%%%%%%%%%%%%%%%%%%%%%%
\begin{eqnarray}
	H 	&=&	\sum_{i \sigma} \epsilon_{i} c_{i\sigma}^{\dag}c_{i\sigma} 
			- t \sum_{<ij> \sigma} c_{i\sigma}^{\dag}c_{j\sigma} 
			+ \epsilon_{d} \sum_{\sigma}n_{d \sigma} 
													\nonumber \\
		& & + V\sum_{\sigma}(d_{\sigma}^{\dag}c_{0\sigma} + H.c.) 
			+ Un_{d\uparrow}n_{d\downarrow}
\end{eqnarray}
%%%%%%%%%%%%%%%%%%%%%%%%%%%%%%%%%%%%%%%%%%%%%%%%%%%%%%%%%%%%%%%%
where random on-site potentials $\epsilon_{i}$ are chosen to be a flat distribution in the interval $[-W, W]$ under the condition $\sum_{i} \epsilon_{i} = 0$, 
$<i,j>$ denotes the summation of the nearest-neighbor sites, and $n_{d \sigma}=d_{\sigma}^{\dag}d_{\sigma}$. 
The system is a $41 \times 41$ square lattice and a magnetic impurity is located at $i=0$. 
For $W \geq 3.0$, the conduction electron is localized with the localization length $\xi(W) \leq 37.5$ \cite{MacKinnon}. Thus, the system for $W \geq 3.0$ is probably in the strongly localized regime at low temperature. 
We set the condition $\epsilon_{d}+(1/2)U=0$. In the case without a random potential, this condition represents the electron-hole symmetry. 
We use the parameters $U=2.0$ and $V=-1.0$ in units of $t$. 
The resonance width can be estimated as $\Delta \equiv \pi \rho_0 V^2 \sim 0.39$, where $\rho_0$ is the density of states at the Fermi energy and can be of the order of the inverse of the bandwidth ($\sim 1/8$). Therefore, the system is in the Kondo regime. The Kondo temperature without randomness can be estimated as \cite{TW}   
$T_K^0 = [V^2 U/\pi^2]^{1/2} \exp [-(\pi/4)(U/V^2 - V^2/U)] \sim 0.14$.

Using the QMC procedure reported in refs. 13 and 14, we calculate the susceptibility of a magnetic impurity, 
%%%%%%%%%%%%%%%%%%%%%%%%%%%%%%%%%%%%%%%%%%%%%%%%%%%%%%%%%%%%%%%%
\begin{eqnarray}
%
%	\chi (T) = \Delta\tau\sum_{l=0}^{L-1}
%	<[n_{d\uparrow}(\tau_{l}) - n_{d\downarrow}(\tau_{l})]
%	[n_{d\uparrow} - n_{d\downarrow}]>,
	\lefteqn{\chi (T) = }			\nonumber \\
	& & \Delta\tau\sum_{l=0}^{L-1}
	\left \langle 
		\left [
			n_{d\uparrow}(\tau_{l}) - n_{d\downarrow}(\tau_{l})
		\right ]
		\left [
			n_{d\uparrow}(\tau_{0}) - n_{d\downarrow}(\tau_{0})
		\right ]
	\right \rangle , 
\end{eqnarray}
%%%%%%%%%%%%%%%%%%%%%%%%%%%%%%%%%%%%%%%%%%%%%%%%%%%%%%%%%%%%%%%%
where $\Delta\tau=\beta/L$ with $\beta=1/(k_{\rm B} T)$ and $L$ is the Trotter number. We set $g\mu_B/2=1$. 
We first numerically diagonalize the Hamiltonian without local correlation effects between $d$ electrons to obtain the initial Green functions. 
We then calculate the local correlation effects by the QMC procedure, and obtain the Green functions needed to calculate physical quantities. 
We use the Trotter time-slice size $\Delta\tau=0.25$. 
An open boundary condition is employed. 
In order to see the effect of the boundary condition on the low-energy properties, we have carried out QMC calculation, applying also the periodic boundary condition. The difference between the two results is negligibly small. 
We have run 50000 Monte Carlo sweeps for measurements after 1000 sweeps in the warming-up run, and have taken an average over the Monte Carlo sweeps.  Note that the effect of a negative sign problem is negligibly small because of the conditions $\epsilon_{d}+(1/2)U=0$ and $\sum_{i} \epsilon_{i} = 0$.

%%%%%%%%%%%%%%%%%%%%%%%%%%%%%%%%%%%%%%%%%%%%%%%%%%%%%%%%%%%%%%%%%%%%%%%%%%%%%
%                        FIGURES 1                                          %
%%%%%%%%%%%%%%%%%%%%%%%%%%%%%%%%%%%%%%%%%%%%%%%%%%%%%%%%%%%%%%%%%%%%%%%%%%%%%
%
\begin{figure}[b]
\includegraphics[trim=2cm 4.5cm 1cm 1cm,clip,width=8cm]{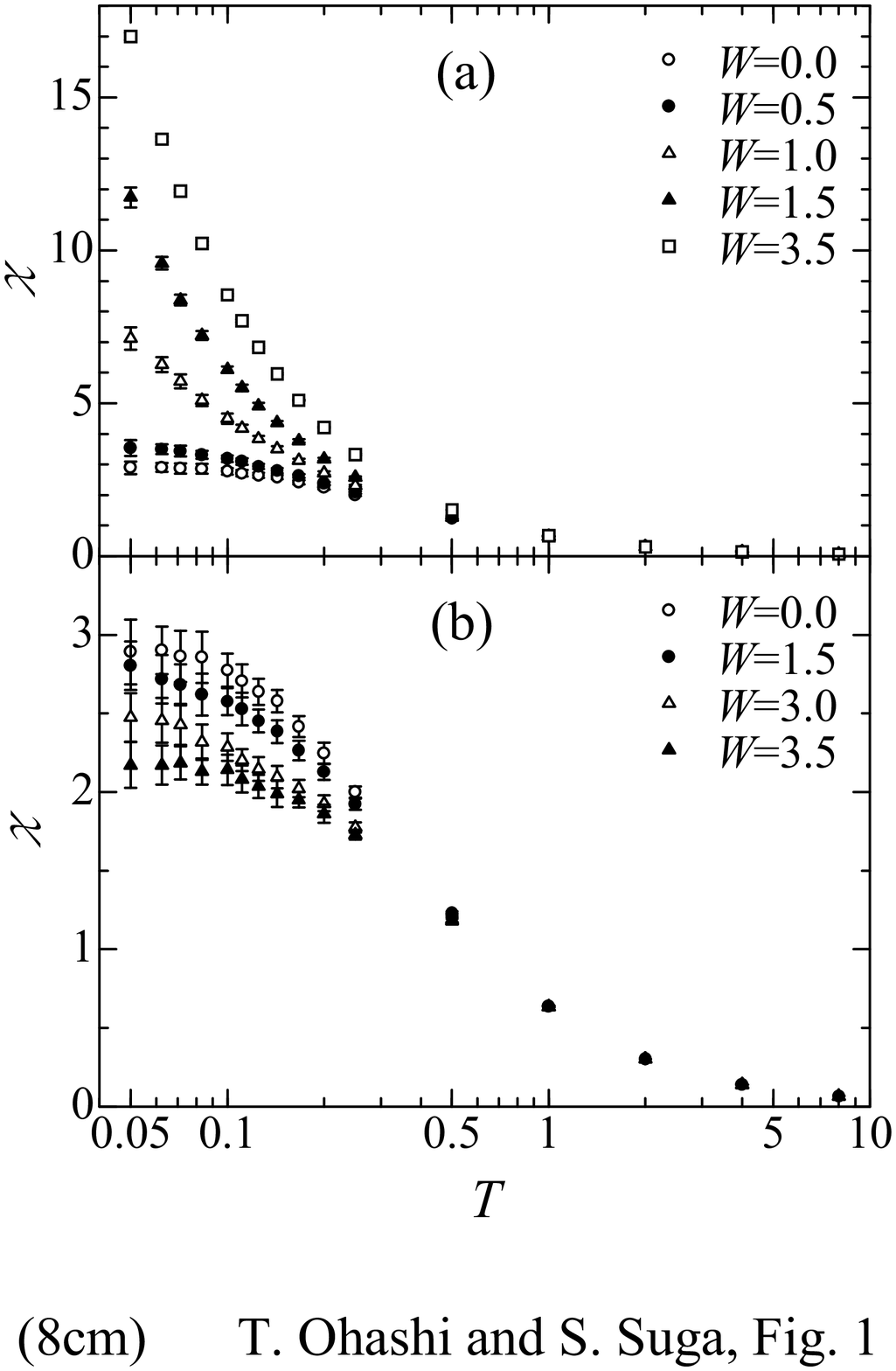}
\caption{
Susceptibility for $0 \leq W \leq 3.5$. A magnetic impurity is located  (a) at the center of the $41 \times 41$ square lattice, and (b) at one of four neighboring sites of the center. Parameters used were $V=1$ and $U=2$ in units of $t$. The Kondo temperature without randomness can be estimated as $T_{K}^0 \sim 0.14$. }
\label{fig:1}
\end{figure}
%
%%%%%%%%%%%%%%%%%%%%%%%%%%%%%%%%%%%%%%%%%%%%%%%%%%%%%%%%%%%%%%%%%%%%%%%%%%%%%
%                        FIGURES 1 END                                      %
%%%%%%%%%%%%%%%%%%%%%%%%%%%%%%%%%%%%%%%%%%%%%%%%%%%%%%%%%%%%%%%%%%%%%%%%%%%%%

In Fig. 1(a), $\chi (T)$ versus $\log T$ is shown as a function of $W$. 
A magnetic impurity is located at the center of the $41 \times 41$ square lattice. In the case without randomness, we observe an ordinary Kondo behavior. 
In this realization of a random potential, the susceptibility of a magnetic impurity shows a divergence for $1.0 \leq W \leq 3.5$ at low temperatures, while it shows a local Fermi-liquid behavior for $W=0.5$. 
Note that we have confirmed that the data in the latter case can be well fitted by $\chi (T) = aT^2 +b$. 
We evaluate the divergent exponent $\alpha$ for $\chi (T) \sim T ^{-\alpha}$ using data below $T_K^0 \sim 0.14$ by the least-squares method. The results are summarized in Table I. 
For $W=1.0$ and $1.5$, the susceptibility of a magnetic impurity diverges with a power weaker than that of a free spin, while for $W \geq 2.0$, a magnetic impurity behaves like a free spin. The results indicate that the impurity spin at this site is not completely screened by the spin of the conduction electron for $W \geq 1.0$. 

%%%%%%%%%%%%%%%%%%%%%%%%%%%%%%%%%%%%%%%%%%%%%%%%%%%%%%%%%%%%%%%%%%%%%%%%%%%%
%                          TABLE  1                                        %
%%%%%%%%%%%%%%%%%%%%%%%%%%%%%%%%%%%%%%%%%%%%%%%%%%%%%%%%%%%%%%%%%%%%%%%%%%%%
% 
\begin{table}[t]
\begin{tabular}{c|cccccc} \hline 
 {$W$} & 
 {1.0} & 
 {1.5} & 
 {2.0} & 
 {2.5} & 
 {3.0} & 
 {3.5} \\ \hline
 $\alpha$ & 
 0.513 & 
 0.938 & 
 0.981 & 
 0.996 & 
 1.000 & 
 0.999 \\
\hline 
\end{tabular}
\caption{
Divergence exponents of $\chi \sim T^{-\alpha}$ in Fig. 1(a) for $1.0 \leq W 	\leq 3.5$. }
\end{table}
%%%%%%%%%%%%%%%%%%%%%%%%%%%%%%%%%%%%%%%%%%%%%%%%%%%%%%%%%%%%%%%%%%%%%%%%%%%%
%                          TABLE  1 END                                    %
%%%%%%%%%%%%%%%%%%%%%%%%%%%%%%%%%%%%%%%%%%%%%%%%%%%%%%%%%%%%%%%%%%%%%%%%%%%%
%
%%%%%%%%%%%%%%%%%%%%%%%%%%%%%%%%%%%%%%%%%%%%%%%%%%%%%%%%%%%%%%%%%%%%%%%%%%%%%
%                        FIGURES  2                                         %
%%%%%%%%%%%%%%%%%%%%%%%%%%%%%%%%%%%%%%%%%%%%%%%%%%%%%%%%%%%%%%%%%%%%%%%%%%%%%
\begin{figure}[b]
\includegraphics[trim=2cm 12cm 1cm 2cm,clip,width=8cm]{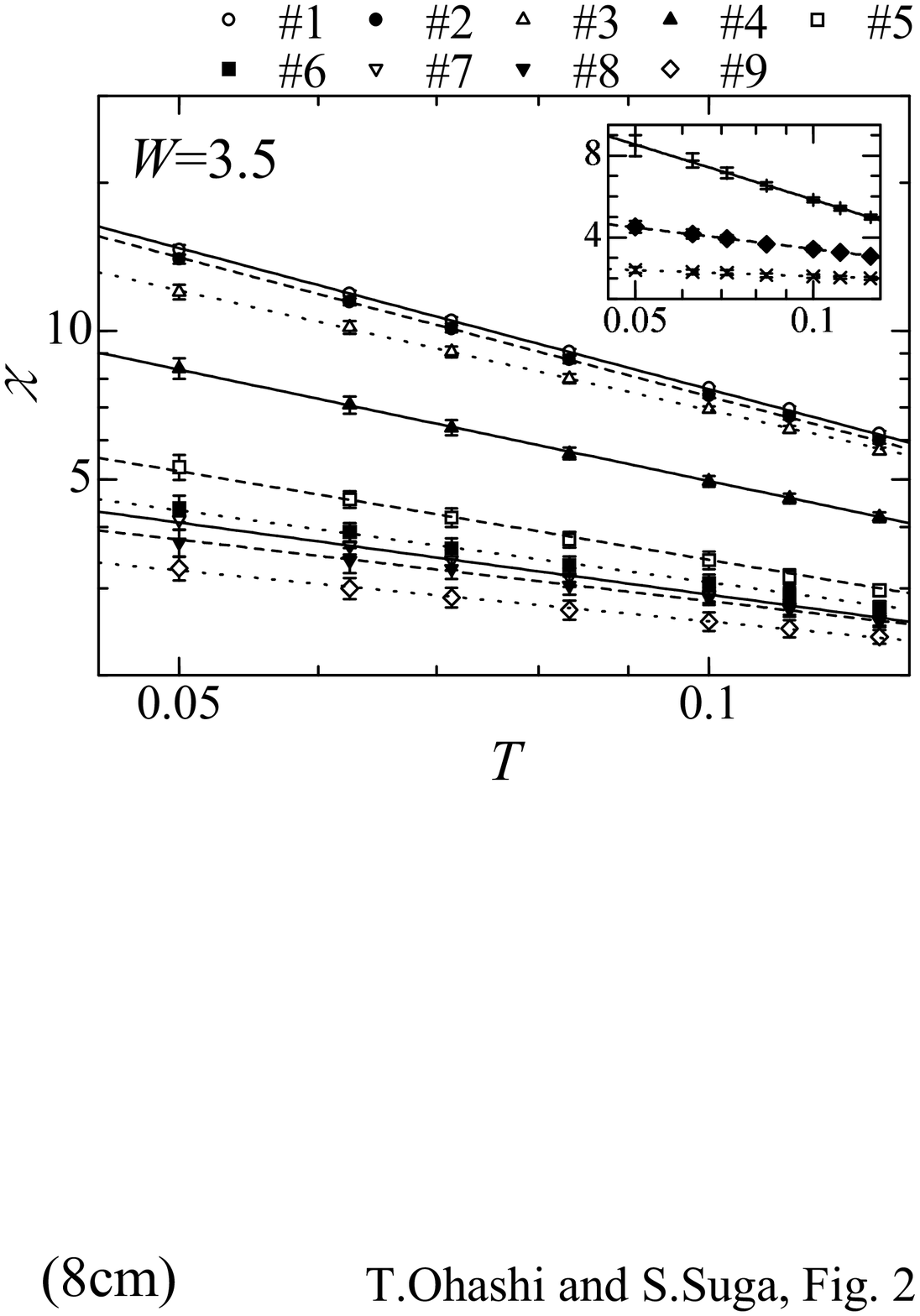}
\caption{
$\log \chi (T)$ versus $\log T$ for weak power-law divergence for $W=3.5$. A magnetic impurity in each result is located at the different position around the center. Parameters and the random potential realized were the same as those in Fig. 1. Solid, broken and dotted lines are fitted by the least-squares method. 
Inset: $\chi (T)$ versus $\log T$ for logarithmic divergence. 
}
\label{fig:2}
\end{figure}
%%%%%%%%%%%%%%%%%%%%%%%%%%%%%%%%%%%%%%%%%%%%%%%%%%%%%%%%%%%%%%%%%%%%%%%%%%%%%
%                        FIGURES  2 END                                     %
%%%%%%%%%%%%%%%%%%%%%%%%%%%%%%%%%%%%%%%%%%%%%%%%%%%%%%%%%%%%%%%%%%%%%%%%%%%%%

Shifting the position of a magnetic impurity twenty-four times around the center in the same realization of the random potential, we calculate the susceptibility of a magnetic impurity. 
At seven positions, the susceptibility shows a local Fermi-liquid behavior irrespective of $W$, while at the remaining seventeen positions, the susceptibility shows a divergence behavior with increasing $W$. 
In the former case, a typical example is shown in Fig. 1(b). 
In contrast to the results in Fig. 1(a), the susceptibility at this site shows a local Fermi-liquid behavior for $0 \leq W \leq 3.5$ with monotonic decrease in $\chi (0)$ as a function of $W$. 
In the latter case, a logarithmic divergence can be seen at three positions, as well as a weak power-law divergence at nine positions. 
The results for weak power-law and logarithmic divergences are depicted in Fig. 2 for $W=3.5$. The divergent exponents are evaluated in the same way as in Fig. 1(a) and are presented in Table II. 
In Table III, the behavior of $\chi$ and $\epsilon_{i}$ for $W=3.5$ 
%
%%%%%%%%%%%%%%%%%%%%%%%%%%%%%%%%%%%%%%%%%%%%%%%%%%%%%%%%%%%%%%%%%%%%%%%%%%%%
%                          TABLE  2                                        %
%%%%%%%%%%%%%%%%%%%%%%%%%%%%%%%%%%%%%%%%%%%%%%%%%%%%%%%%%%%%%%%%%%%%%%%%%%%%
% 
\begin{table}[t]
\begin{tabular}{c|ccccccccc} \hline 
%Result number 
& 
$\# 1$ & 
$\# 2$ & 
$\# 3$ & 
$\# 4$ & 
$\# 5$ & 
$\# 6$ & 
$\# 7$ & 
$\# 8$ & 
$\# 9$ \\ \hline
$\alpha$ & 
0.952 & 
0.938 & 
0.807 & 
0.751 & 
0.597 & 
0.485 & 
0.485 & 
0.414 & 
0.343 \\
\hline 
\end{tabular}
\caption{
Weak divergence exponents of $\chi \sim T^{-\alpha}$ in Fig. 2. }
\end{table}
%%%%%%%%%%%%%%%%%%%%%%%%%%%%%%%%%%%%%%%%%%%%%%%%%%%%%%%%%%%%%%%%%%%%%%%%%%%%
%                          TABLE  2 END                                    %
%%%%%%%%%%%%%%%%%%%%%%%%%%%%%%%%%%%%%%%%%%%%%%%%%%%%%%%%%%%%%%%%%%%%%%%%%%%%
%
%%%%%%%%%%%%%%%%%%%%%%%%%%%%%%%%%%%%%%%%%%%%%%%%%%%%%%%%%%%%%%%%%%%%%%%%%%%%
%                          TABLE  3                                        %
%%%%%%%%%%%%%%%%%%%%%%%%%%%%%%%%%%%%%%%%%%%%%%%%%%%%%%%%%%%%%%%%%%%%%%%%%%%%
% 
\begin{table}[h]
\begin{tabular}{|c||c|c|c|c|c|}
\hline
& 19 & 20 & 21 & 22 & 23 \\
\hline\hline
19 & FL	& \#7 &	FL & FL & \#3 \\
& -0.670 & -0.844 & -2.764 & -0.580 & -2.326 \\
\hline
20 & FS	& FS & FL & log-D & FL \\
& 2.533	& 2.116	& -0.642 & 2.404 & 1.787 \\
\hline
21 & \#8 & \#5 & FS & \#1 & FL \\
& -0.102 & 0.108 & 3.054 & -2.342 & 1.407 \\
\hline
22 & FS	& \#4 &	FL & FS	& FS \\
& -2.177 & 1.336 & 0.013 & 2.364 & -2.667 \\
\hline
23 & log-D & \#6 & log-D & \#9 & \#2 \\
& 0.094	& -0.454 & -0.099 & 0.454 & 2.985 \\
\hline
\end{tabular}
\caption{
Behavior of $\chi$ and $\epsilon_{i}$  for $W=3.5$ at each shifted position of a magnetic impurity. The average $\epsilon_{i}$ is zero. The numbers in the top row and the column of the left-hand side represent the coordinates along the horizontal and vertical axes, respectively. The position (21, 21) is the center of the system. FL, log-D and FS denote local Fermi-liquid behavior, logarithmic divergence and behavior of a free spin, respectively. $\# 1 \sim \# 9$ correspond to the results in Fig. 2. }
\end{table}
%
%%%%%%%%%%%%%%%%%%%%%%%%%%%%%%%%%%%%%%%%%%%%%%%%%%%%%%%%%%%%%%%%%%%%%%%%%%%%
%                          TABLE  3 END                                    %
%%%%%%%%%%%%%%%%%%%%%%%%%%%%%%%%%%%%%%%%%%%%%%%%%%%%%%%%%%%%%%%%%%%%%%%%%%%%
%
%%%%%%%%%%%%%%%%%%%%%%%%%%%%%%%%%%%%%%%%%%%%%%%%%%%%%%%%%%%%%%%%%%%%%%%%%%%%%
%                        FIGURES  3                                         %
%%%%%%%%%%%%%%%%%%%%%%%%%%%%%%%%%%%%%%%%%%%%%%%%%%%%%%%%%%%%%%%%%%%%%%%%%%%%%
%
\begin{figure}[b]
\includegraphics[trim=2cm 10cm 0cm 5cm,clip,width=8cm]{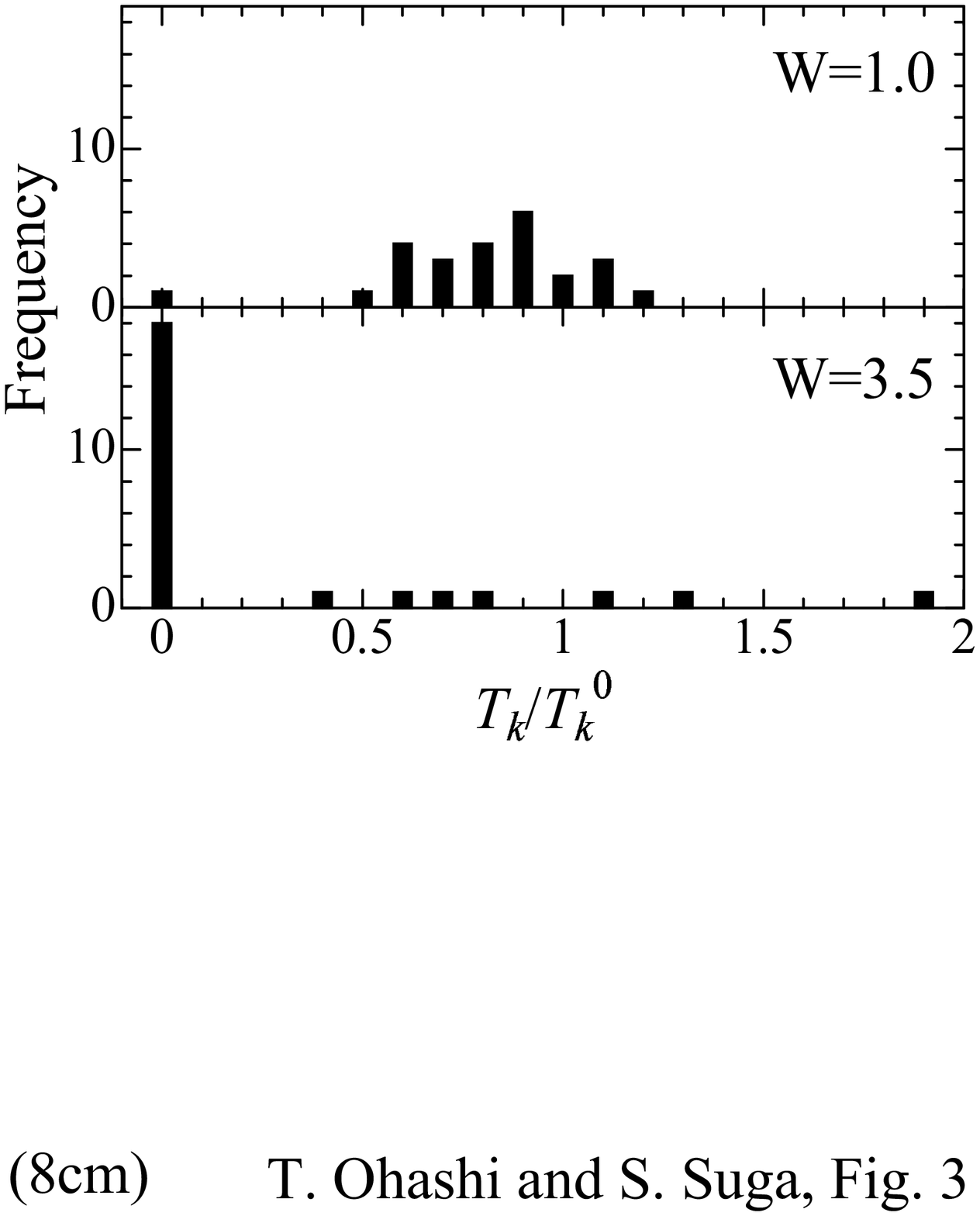}
\caption{
Distribution of the Kondo temperature for $W=1.0$ and $3.5$. The position of a magnetic impurity is shifted twenty-four times around the center. $T_{K}^0$ is the Kondo temperature without randomness. 
}
\label{fig:3}
\end{figure}
%
%%%%%%%%%%%%%%%%%%%%%%%%%%%%%%%%%%%%%%%%%%%%%%%%%%%%%%%%%%%%%%%%%%%%%%%%%%%%%
%                        FIGURES  3 END                                     %
%%%%%%%%%%%%%%%%%%%%%%%%%%%%%%%%%%%%%%%%%%%%%%%%%%%%%%%%%%%%%%%%%%%%%%%%%%%%%
%
at each position of a magnetic impurity is summarized. 
From the results, we conclude that the local moment in 2D disordered electron systems is screened or unscreened by the spin of the conduction electron, depending on the position of a magnetic impurity. 
Since $\chi(0) = (2\pi T_K)^{-1}$ with $T_K$ being the Kondo temperature \cite{TW}, this conclusion indicates that the Kondo temperature has a spatial distribution down to $T_K=0$. 

On the basis of the results for the susceptibility of a magnetic impurity, we estimate the distribution of the Kondo temperature. 
When the susceptibility shows a Fermi-liquid behavior, we extrapolate $\chi (0)$ by the least-squares method using $\chi (T) = aT^2 +b$.  
The results are summarized in Fig. 3. 
At $W=1.0$, the Kondo temperature is distributed around $T_K=0.9 \, T_K^0$, where the largest weight exists, and a small weight lies at $T_K=0$. At $W=3.5$, the distribution of $T_K$ becomes wider and the weight at $T_K=0$ increases considerably with reduced weights for $T_K \neq 0$. 
We have shown that the spatial distribution of the Kondo temperature can be caused only by the effects of a random potential. 

%%%%%%%%%%%%%%%%%%%%%%%%%%%%%%%%%%%%%%%%%%%%%%%%%%%%%%%%%%%%%%%%%%%%%%%%%%%%%
%                        FIGURES  4                                         %
%%%%%%%%%%%%%%%%%%%%%%%%%%%%%%%%%%%%%%%%%%%%%%%%%%%%%%%%%%%%%%%%%%%%%%%%%%%%%
%
\begin{figure}[t]
\includegraphics[trim=1cm 14cm 0cm 1cm,clip,width=8cm]{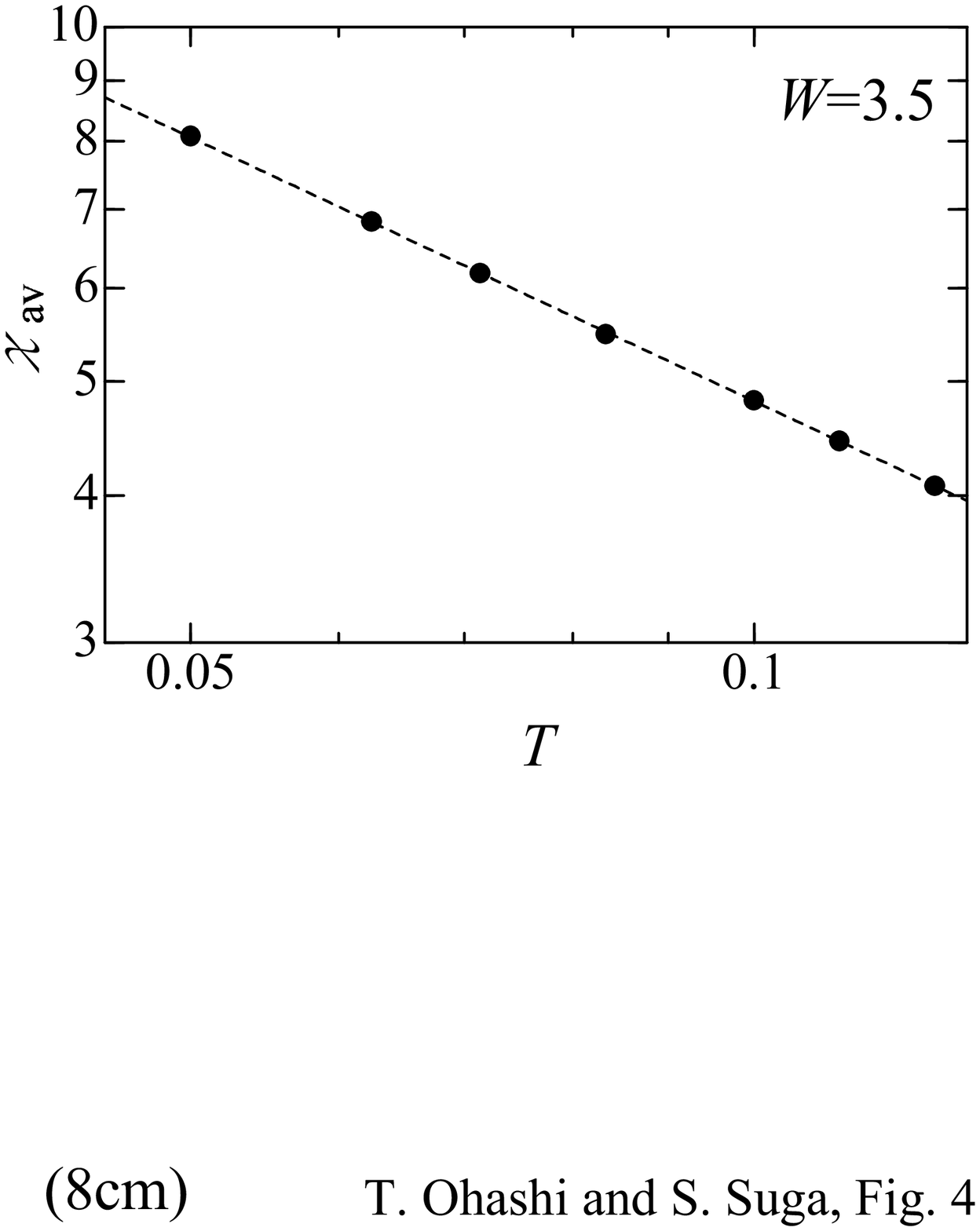}
\caption{
Average susceptibility for $W = 3.5$ in $0.05 < T < T_{K}^0 \sim 0.14$. 
The broken line is fitted by the least-squares method: 
$\chi_{\rm av}=0.867 \, T^{-0.744}$.
}
\label{fig:4}
\end{figure}
%
%%%%%%%%%%%%%%%%%%%%%%%%%%%%%%%%%%%%%%%%%%%%%%%%%%%%%%%%%%%%%%%%%%%%%%%%%%%%%
%                        FIGURES  4 END                                     %
%%%%%%%%%%%%%%%%%%%%%%%%%%%%%%%%%%%%%%%%%%%%%%%%%%%%%%%%%%%%%%%%%%%%%%%%%%%%%

In the disordered system, we take an ensemble average over the distribution of random potentials in order to extract the universal properties. Judging from our numerical results, it seems that the susceptibility of a magnetic impurity does not show the universal behavior at low temperature. 
Therefore, in 2D strongly disordered electron systems with dilute magnetic impurities, where each magnetic impurity acts as a single magnetic impurity, the observable susceptibility at low temperature may be obtained as the arithmetic mean over the susceptibility of each magnetic impurity. The average susceptibility thus obtained in the random potential for $W=3.5$ realized in Fig. 1 is shown in Fig. 4. The average susceptibility for $0.05 <T < T_{K}^0$ shows a weak power-law divergence $\chi_{\rm av} \sim T^{-\alpha}$ with $\alpha \sim 0.744$, indicating a non-Fermi-liquid behavior. 

To obtain local information about Kondo screening, we calculate the correlation function between the local moment at the $0$ site and the spin of the conduction electron at the $i$ site,  
%%%%%%%%%%%%%%%%%%%%%%%%%%%%%%%%%%%%%%%%%%%%%%%%%%%%%%%%%%%%%%%%
\begin{eqnarray}
<S^z_0 s^z_i> =  \frac{1}{4}<[n_{d\uparrow} - n_{d\downarrow}] 
                  [n_{i\uparrow} - n_{i\downarrow}]> , 
\end{eqnarray}
%%%%%%%%%%%%%%%%%%%%%%%%%%%%%%%%%%%%%%%%%%%%%%%%%%%%%%%%%%%%%%%%
where $n_{i\sigma} = c_{i\sigma}^{\dag}c_{i\sigma}$. 
In Figs. 5(a) and 5(b), we show the spin correlation functions along the vertical and horizontal axes of the square lattice, respectively.  
The realized random potential and the position of the local moment are the same as those used in Figs. 1(a) and 1(b). Parameters used were $W=3.0$ and $\beta=20$. 
Thus, the results shown in Fig. 1(a) (open circles) correspond to the unscreened case, while those shown in Fig. 1(b) (full triangles)  correspond to the screened case. In the former case, the antiferromagnetic correlation at the magnetic impurity is suppressed due to randomness. 
The suppression of the antiferromagnetic correlation at the magnetic impurity is commonly seen in other locations of a magnetic impurity where the susceptibility shows a divergence. 
This suppression probably causes the incomplete screening. 
Although the antiferromagnetic correlation at the magnetic impurity is not suppressed in the screened case, the behavior of the spin correlation function around a magnetic impurity is different from that for $W=0$; the ferromagnetic correlation is reduced. 

%%%%%%%%%%%%%%%%%%%%%%%%%%%%%%%%%%%%%%%%%%%%%%%%%%%%%%%%%%%%%%%%%%%%%%%%%%%%%
%                        FIGURES  5                                         %
%%%%%%%%%%%%%%%%%%%%%%%%%%%%%%%%%%%%%%%%%%%%%%%%%%%%%%%%%%%%%%%%%%%%%%%%%%%%%
%
\begin{figure}[t]
\includegraphics[trim=1cm 2.5cm 1cm 0cm,clip,width=8cm]{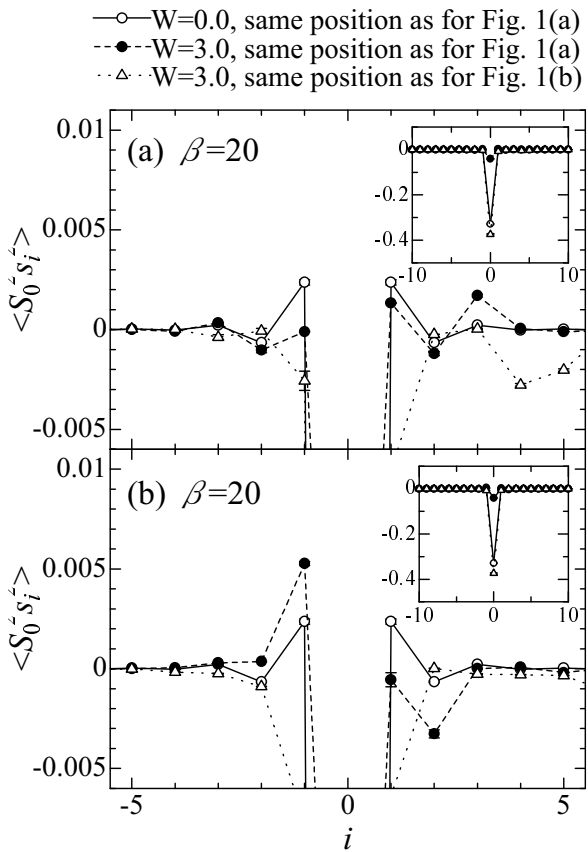}
\caption{
Correlation function $<S^z_0 s^z_i>$ along the (a) vertical and (b) horizontal axes of the square lattice. Random potentials and the position of the local moment are the same as those used in Figs. 1(a) and 1(b). Parameters used were $V=1$, $U=2$, $W=3.0$, and $\beta=20$. 
Inset: $<S^z_0 s^z_i>$ in a wide scope. 
}
\label{fig:5}
\end{figure}
%
%%%%%%%%%%%%%%%%%%%%%%%%%%%%%%%%%%%%%%%%%%%%%%%%%%%%%%%%%%%%%%%%%%%%%%%%%%%%%
%                        FIGURES  5 END                                     %
%%%%%%%%%%%%%%%%%%%%%%%%%%%%%%%%%%%%%%%%%%%%%%%%%%%%%%%%%%%%%%%%%%%%%%%%%%%%%

In summary, we have investigated the Kondo effect in 2D disordered electron systems using a finite-temperature QMC method. We have shown that depending on the position of a magnetic impurity, its local moment can be screened or unscreened by the spin of the conduction electron. 
In the latter case, the antiferromagnetic spin correlation between the local moment and the conduction electron is suppressed around a magnetic impurity. 
We have demonstrated that the effects of randomness induce a distribution of the Kondo temperature down to $T_K=0$, and that the average susceptibility diverges with a weak power law at low temperature in strongly disordered systems. 

We would like to thank K. Naito and T. Saso for useful comments and valuable discussions. 
Numerical computation was partly carried out at the Supercomputer Center, the Institute for Solid State Physics, University of Tokyo. 
This work was partly supported by a Grant-in-Aid for Scientific Research from the Ministry of Education, Culture, Sports, Science and Technology, Japan. 

%%%%%%%%%%%%%%%%%%%%%%%%%%%%%%%%%%%%%%%%%%%%%%%%%%%%%%%%%%%%%%%%%%%%%%

\end{document}